# Observation of superconductivity-induced leading-edge gap in Sr-doped La$_3$Ni$_2$O$_7$ thin films

Wenjie Sun[1,2,†], Zhicheng Jiang[3,†], Bo Hao[1,2,†], Shengjun Yan[1,2], Hongyi Zhang[1,2], Maosen Wang[1,2], Yang Yang[4], Haoying Sun[1,2], Zhengtai Liu[5], Dianxiang Ji[4], Zhengbin Gu[1,2], Jian Zhou[1,2,*], Dawei Shen[3,*], Donglai Feng[6,*], Yuefeng Nie[1,2,7,*]

[1]National Laboratory of Solid State Microstructures, Jiangsu Key Laboratory of Artificial Functional Materials, College of Engineering and Applied Sciences, Nanjing University, Nanjing, 210093, China.

[2]Collaborative Innovation Center of Advanced Microstructures, Nanjing University, Nanjing, 210093, China.

[3]National Synchrotron Radiation Laboratory and School of Nuclear Science and Technology, University of Science and Technology of China, Hefei, 230026, China.

[4]Department of Applied Physics, The Hong Kong Polytechnic University, Hung Hom, Hong Kong, 999077, China.

[5]Shanghai Synchrotron Radiation Facility, Shanghai Advanced Research Institute, Chinese Academy of Sciences, Shanghai, 201210, China.

[6]New Cornerstone Science Laboratory, Hefei National Laboratory, Hefei, 230088, China

[7]Jiangsu Physical Science Research Center, Nanjing, 210093, China.

[†]These authors contributed equally to this work.

[*]Correspondence to: zhoujian@nju.edu.cn (J.Z.); dwshen@ustc.edu.cn (D.S.); dlfeng@ustc.edu.cn (D.F.); ynie@nju.edu.cn (Y.N.)

**Abstract**

The discovery of high-temperature superconductivity in pressurized bulk La$_3$Ni$_2$O$_7$ has ignited significant interest in nickelate superconductors[1-3]. Unlike cuprates, where superconductivity predominantly originates from the $3d_{x^2-y^2}$ orbital, nickelates exhibit additional complexities involving contributions from the $3d_{z^2}$ orbital, prompting fundamental questions about their pairing mechanisms. Despite recent progress in stabilizing superconductivity in La$_3$Ni$_2$O$_7$ thin films at ambient pressure[4,5], direct spectroscopic evidence of the superconducting gap opening remains elusive. Here, we present an *in-situ* angle-resolved photoemission spectroscopy study of Sr-doped superconducting La$_3$Ni$_2$O$_7$ thin films. Fermi surface mapping reveals Ni-$3d_{x^2-y^2}$-derived $\alpha$ and $\beta$ pockets, with orbital fillings of 0.11±0.02 electrons and 0.66±0.03 holes per Ni, respectively, resulting in a total of 0.45±0.04 electrons for each Ni. These bands exhibit moderate electron correlations, characterized by a band renormalization factor of 3-4. Notably, both $\alpha$ and $\beta$ bands exhibit leading-edge shifts across the superconducting transition, with gap magnitude of ~1-2 meV at Fermi momenta along the Brillouin zone diagonal and slightly away from the zone diagonal, deviating from the conventional $d_{x^2-y^2}$-wave gap structure. Additionally, the Ni-$3d_{z^2}$-derived $\gamma$ band lies ~75 meV below the Fermi level, indicating a $3d_{x^2-y^2}$-dominated fermiology in this compound.

**Main text**

**Introduction**

The discovery of superconducting bilayer and trilayer nickelates under high pressure marks a significant milestone in strongly correlated electron system, providing a unique platform to explore the microscopic origins of high-temperature (high-$T_c$) superconductivity[1-3,6,7]. Unlike cuprates, bilayer nickelates exhibit $d^{7.5}$ electron filling and multiband electronic structures. Consequently, various superconducting pairing symmetries have been theoretically proposed, reflecting various relative contributions of Ni-$3d_{x^2-y^2}$ and Ni-$3d_{z^2}$ orbitals[8-19]. Additionally, the $\gamma$ band, characterized by a large density of state with predominate Ni-$3d_{z^2}$ contributions, is predicted to cross the Fermi level ($E_F$) under high pressure[8,9,12,16-19], potentially playing a crucial role in superconductivity. This differs significantly from high-$T_c$ cuprates, where the Cu-$3d_{x^2-y^2}$ orbital plays a dominant role in Cooper pairing. Understanding these distinctions, particularly the superconducting pairing mechanism between the bilayer nickelates and the cuprates requires momentum-resolved electronic structure measurements in the superconducting states. However, such measurements are technically challenging, as stabilizing the bulk superconducting phase necessitates high-pressure conditions that severely limit the applicability of surface-sensitive experimental probes like angle-resolved photoemission spectroscopy (ARPES).

This challenge has recently been circumvented through the stabilization of ambient-pressure superconductivity in compressively strained La$_3$Ni$_2$O$_7$ thin films grown on SrLaAlO$_4$ substrate[4,5,20,21]. However, the sensitivity to oxygen vacancies and the high-pressure post-growth ozone annealing make it extremely difficult to optimize superconducting properties and surface quality simultaneously, complicating subsequent ARPES measurements on superconducting films. Probably for this reason, previous ARPES studies have produced conflicting conclusions[22,23]. For instance, one study reported that the $\gamma$ band crosses $E_F$ in thin films[22], whereas another found this band is still located approximately 70 meV below $E_F$ in both as-grown non-superconducting and superconducting films[23]. Moreover, laser-based ARPES measurements identified a persistent nodeless energy gap extending significantly above the superconducting transition temperature, the origin of which remains unresolved[24]. Therefore, further experimental investigations into the electronic structure, particularly its evolution across the superconducting transition, are urgently required.

In this work, we investigate the electronic structure of Sr-doped superconducting La₃Ni₂O₇ thin film using synchrotron-based ARPES and density functional theory (DFT) calculations. The Fermi surface (FS) primarily consists of an electron pocket (α) centered at (0,0) at the Brillouin zone (BZ), and a larger hole pocket (β) centered at (π, π), consistent with theoretical calculations. Compared to the DFT bare bands, both the α and β bands show moderate band renormalizations, albeit slightly larger than those in single crystals[25]. Furthermore, the Ni-$3d_{z^2}$-derived γ band is located approximately 75 meV below $E_F$, contrasting with the predicted energy shift toward $E_F$ in superconducting bulk samples under high-pressure condition[8]. Notably, a leading-edge shift of around 2 meV is observed for both α and β bands upon cooling through the superconducting transition temperature, implicating that Ni-$3d_{x^2-y^2}$ orbitals are essential for the Cooper pairing. These results provide critical experimental insights addressing the ongoing debate regarding the multiband superconductivity mechanism in the bilayer nickelates.

**Fermi surface topology of La$_{2.79}$Sr$_{0.21}$Ni$_2$O$_7$**

La$_{2.79}$Sr$_{0.21}$Ni$_2$O$_7$ thin films were grown on SrLaAlO$_4$ substrates via oxide molecular beam epitaxy (MBE)[26]. The film thickness is kept as 2 unit cell (u.c.) throughout this paper unless otherwise mentioned. Before ARPES measurements, thin films were *in-situ* ozone annealed in the MBE chamber, to minimize oxygen vacancies and optimize superconducting properties (see Methods for details). The annealed thin films exhibit a clear superconducting transition with the onset temperature of 37 K, as demonstrated by representative measurements (Fig. 1**a**). Scanning transmission electron microscopy (STEM) and x-ray diffraction confirm a uniform bilayer structure without detectable intergrowth of other Ruddlesden–Popper (RP) phases (Fig. 1**b-c** and Extended Data Fig. 1). Atomic force microscopy measurements confirm atomically flat surfaces of the superconducting thin films (Fig. 1**d**).

The FS mappings are taken at 7 K, as shown in Fig. 2**a-c**. There are mainly two bands that cross $E_F$, which corresponds well with the theoretical calculations and previous experimental results. Among them, the $\beta$ pocket is clearly revealed using 45- and 120-eV photons, while the $\alpha$ pocket is more prominent under 90-eV photons, especially in the second BZ. It is noted that the $\gamma$ band is absent in the FS mapping, suggesting that it is below $E_F$ (as discussed later), which is consistent with the previous ARPES results in superconducting $La_2PrNi_2O_7$ films[23]. The experimentally derived $\alpha$ and $\beta$ pockets agree with the calculated FS for $La_3Ni_2O_7$ under 2% compressive strain (Fig. 2**d**). The absence of obvious hole-doping effect by Sr substitution may be explained by the variation of oxygen contents, which can compensate the hole doping, given the delicate oxygen contents in this system. According to the Luttinger theorem, the orbital fillings for $\alpha$ and $\beta$ bands are estimated to be 0.11±0.02 electron/Ni and 0.66±0.03 hole/Ni, respectively, which corresponds to a total filling of 0.45±0.04 electron/Ni, close to that in single crystals[25] and larger than that in $La_{2.85}Pr_{0.15}Ni_2O_7$ films[22].

**Band dispersions and renormalizations**

The representative energy-momentum cuts are measured along high-symmetric directions using different photon energies, as shown in Fig 3. Clear band renormalizations can be observed for $\alpha$ and $\beta$ bands when compared with our calculated results shown in Fig. 3**a-d**. The renormalization factor for Ni-$3d_{x^2-y^2}$-derived $\alpha$ and $\beta$ bands is 3-4, which is in agreement with that in superconducting $La_2PrNi_2O_7$ films[23], and is slightly larger than that in non-superconducting bulk crystals[25]. It is also noted that the renormalization factors in $La_{2.79}Sr_{0.21}Ni_2O_7$ thin film exceed those observed in infinite-layer $La_{0.8}Sr_{0.21}NiO_2$ as well[27], indicating an enhanced electron correlations in bilayer nickelates. The waterfall-like dispersion is observed at higher binding energy (above 100 meV), the origin of which remains to be explored[28,29].

As a major distinction from cuprates, $γ$ band is believed to be vital in superconducting pairing in bilayer nickelates[8,9,12,16-19]. It is theoretically expected to cross $E_F$ under hydrostatic pressure in bulk crystals[8,30], while moving further away from $E_F$ under 2% biaxial compressive strain[31-33]. In our calculations, it resides ~120 meV below $E_F$ ($U$ = 0 eV), and further dives below $E_F$ upon considering finite $U$ of 4 eV (Extended Data Fig. 2). Experimentally, the band-top of the $γ$ band is located at 75 ± 5 meV below $E_F$, which is clearer in the background-subtracted spectrum (Fig. 3**e-f**). The background is generated by averaging EDCs at the momentum range of 1.5-1.7 1/Å where low energy bands are absent (Extended Data Fig. 3). The downshift of the $γ$ band compared with bulk crystals[8] is qualitatively captured by our DFT calculations, while the quantitative difference can possibly result from the strong electron correlations for this band. We also notice two conflicting photoemission studies of $La_{3-x}Pr_xNi_2O_7$ films regarding the dispersion of $γ$ band[22,23], which may be attributed to different Pr contents and photon-energy-dependent matrix element effects.

**Leading-edge shifts measurements**

To gain deeper insights into the electronic structure evolution across the superconducting transition, we have also performed temperature-dependent ARPES measurements near the diagonal direction of the first BZ where the band dispersion is sharp enough for detailed analysis (Fig. 4 and Extended Data Fig. 4). To optimize the energy resolution, the photon energy was set to 45 eV, with a corresponding measured energy resolution of ~5 meV on a polycrystalline gold reference (Extended Data Fig. 5). We mainly focus on the temperature evolution of $α$ and $β$ bands that cross $E_F$. A momentum cut offset from the BZ diagonal is selected to simultaneously capture and distinguish these two bands. Crucially, a spectral-weight depletion at $E_F$ emerges at low temperature in symmetrized energy distribution curves (EDCs) at individual Fermi momentum ($k_F$) (Fig. 4**e** and **i**, and Extended Data Fig. 6). This gap-like feature undergoes a clear restoration when the temperature exceeds the superconducting onset temperature (Fig. 4**a**). Given the particle-hole symmetry of the features near $E_F$ in EDCs after divided by Fermi-Dirac distribution (Extended Data Fig. 6) and the coincidence between the spectral-weight depletion and $T_c$, the gap is most likely related to the superconducting transition. We also note that this spectral-weight depletion is due to intrinsic effect instead of sample aging, which has been verified by the temperature cycling measurements (Extended Data Fig. 7).

Further quantitative analysis reveals the leading-edge shifts of ~2 meV for both $α$ and $β$ bands, by comparing high-and low-temperature EDCs at individual $k_F$ (Fig. 4**b-d** and **f-h**). Note that although the exact value of leading-edge shifts deviates slightly with the EDC normalization process, the temperature evolution observed here is robust (Fig. 4**d** and **h**, and Extended Data Fig. 8). This temperature evolution can be simply described by the BCS energy-gap function, which yields $T_c$ and "gap-size" of ~40 K and ~2 meV, respectively. As shown in Extended Data Fig. 9, similar leading-edge shifts are observed from symmetric Fermi momenta along the same cut (Fig. 5**f**).

In addition, a leading-edge shift of ~1 meV is also observed along the diagonal direction of the first BZ (Fig. 5**a-e**). The observed leading-edge gap thus deviates from the conventional *d*-wave gap structure observed in the cuprates. Note that the leading-edge shift analysis on EDCs extracted from another symmetric Fermi momentum along the same cut (Fig. 5**f**) yields comparable results (Extended Data Fig. 10), further strengthening our conclusions.

**Discussion**

Given the orbital character of $α$ and $β$ bands, our experimental results indicate the Ni-$3d_{x^2-y^2}$ orbital plays a dominant role in the superconducting state, in line with recent x-ray absorption measurements on $La_{2.85}Pr_{0.15}Ni_2O_7$ thin films[34]. We also note that in early reports in electron-doped cuprates, the superconducting gap size determined by leading-edge shifts is underestimated by a factor of 2-3[35,36]. Consequently, we anticipate the actual superconducting gap to be significantly larger than these leading-edge shifts values. Although higher-quality thin films exhibiting prominent superconducting peaks in ARPES data collected across whole BZ, are needed to finally pin down the gap structure, our data has put strong constraints on theory (Fig. 5**h**).

Note that we could not fully rule out the influence of disorder on the superconducting gap in our samples. The absence of a distinct superconducting coherence peak in the photoemission spectra suggests disorder-induced suppression of superconductivity, likely caused by oxygen vacancies or structural imperfections in our thin films. While an ideal *d*-wave gap exhibits nodes where the gap vanishes, impurities can introduce finite scattering rates, mixing different momentum states and filling the nodes, leading to a small residual gap[37]. Future studies with higher-quality, less disordered thin films are essential to validate this interpretation.

In summary, the ARPES measurements on superconducting $La_{2.79}Sr_{0.21}Ni_2O_7$ thin films reveal a multiband Fermi surface topology that matches the DFT calculations. Among them, the Ni- $3d_{x^2-y^2}$ -derived *α* and *β* bands display obvious band renormalization effects, suggesting moderate electron correlations are intrinsic to the bilayer nickelate system. Unlike *α* and *β* bands that cross $E_F$, *γ* band is located at ~75 meV below $E_F$, which settles the long-standing debate over the Fermi surface topology of this compound. Furthermore, the leading-edge gaps of 1~2 meV have been observed for both *α* and *β* bands, which are located both at and near the "nodal" point, deviating from the conventional *d*-wave symmetry in cuprates and indicating the key role of Ni-$3d_{x^2-y^2}$ orbitals in participating superconducting condensation. Our work provides the spectroscopic evidence for superconductivity-induced leading-edge gaps in the bilayer nickelate system, providing essential guidelines for establishing the minimal model for elucidating the pairing mechanism in this system.

# References

1. Sun, H. *et al.* Signatures of superconductivity near 80 K in a nickelate under high pressure. *Nature* **621**, 493-498 (2023).

2. Hou, J. *et al.* Emergence of high-temperature superconducting phase in pressurized $La_3Ni_2O_7$ crystals. *Chin. Phys. Lett.* **40**, 117302 (2023).

3. Zhang, Y. *et al.* High-temperature superconductivity with zero resistance and strange-metal behaviour in $La_3Ni_2O_{7-\delta}$. *Nat. Phys.* **20**, 1269-1273 (2024).

4. Ko, E. K. *et al.* Signatures of ambient pressure superconductivity in thin film $La_3Ni_2O_7$. *Nature* **638**, 935-940 (2025).

5. Zhou, G. *et al.* Ambient-pressure superconductivity onset above 40 K in $(La,Pr)_3Ni_2O_7$ films. *Nature* **640**, 641-646 (2025).

6. Wang, N. *et al.* Bulk high-temperature superconductivity in pressurized tetragonal $La_2PrNi_2O_7$. *Nature* **634**, 579-584 (2024).

7. Zhu, Y. *et al.* Superconductivity in pressurized trilayer $La_4Ni_3O_{10-\delta}$ single crystals. *Nature* **631**, 531-536 (2024).

8. Luo, Z., Hu, X., Wang, M., Wú, W. & Yao, D.-X. Bilayer two-orbital model of $La_3Ni_2O_7$ under pressure. *Phys. Rev. Lett.* **131**, 126001 (2023).

9. Yang, Q.-G., Wang, D. & Wang, Q.-H. Possible $s_{\pm}$-wave superconductivity in $La_3Ni_2O_7$. *Phys. Rev. B* **108**, L140505 (2023).

10. Lu, C., Pan, Z., Yang, F. & Wu, C. Interplay of two $E_g$ orbitals in superconducting $La_3Ni_2O_7$ under pressure. *Phys. Rev. B* **110**, 094509 (2024).

11. Xia, C., Liu, H., Zhou, S. & Chen, H. Sensitive dependence of pairing symmetry on Ni-$e_g$ crystal field splitting in the nickelate superconductor $La_3Ni_2O_7$. *Nat. Commun.* **16**, 1054 (2025).

12. Lechermann, F., Gondolf, J., Bötzel, S. & Eremin, I. M. Electronic correlations and superconducting instability in $La_3Ni_2O_7$ under high pressure. Preprint at http://arxiv.org/abs/2306.05121 (2023).

13. Fan, Z. *et al.* Superconductivity in nickelate and cuprate superconductors with strong bilayer coupling. *Phys. Rev. B* **110**, 024514 (2024).

14. Gu, Y., Le, C., Yang, Z., Wu, X. & Hu, J. Effective model and pairing tendency in the bilayer Ni-based superconductor $La_3Ni_2O_7$. *Phys. Rev. B* **111**, 174506 (2025).

15. Jiang, K., Wang, Z. & Zhang, F.-C. High-temperature superconductivity in $La_3Ni_2O_7$. *Chin. Phys. Lett.* **41** (2024).

16. Shen, Y., Qin, M. & Zhang, G.-M. Effective bi-layer model Hamiltonian and density-matrix renormalization group study for the high-$T_c$ superconductivity in $La_3Ni_2O_7$ under high pressure. Preprint at http://arxiv.org/abs/2306.07837 (2023).

17. Christiansson, V., Petocchi, F. & Werner, P. Correlated electronic structure of $La_3Ni_2O_7$ under pressure. *Phys. Rev. Lett.* **131**, 206501 (2023).

18. Liu, Y.-B., Mei, J.-W., Ye, F., Chen, W.-Q. & Yang, F. $s_{\pm}$-wave pairing and the destructive role of apical-oxygen deficiencies in $La_3Ni_2O_7$ under Pressure. *Phys. Rev. Lett.* **131**, 236002 (2023).

19  Yang, Y.-f., Zhang, G.-M. & Zhang, F.-C. Interlayer valence bonds and two-component theory for high-$T_c$ superconductivity of $La_3Ni_2O_7$ under pressure. *Phys. Rev. B* **108**, L201108 (2023).

20  Liu, Y. *et al.* Superconductivity and normal-state transport in compressively strained $La_2PrNi_2O_7$ thin films. Preprint at http://arxiv.org/abs/2501.08022 (2025).

21  Hao, B. *et al.* Superconductivity and phase diagram in Sr-doped $La_{3-x}Sr_xNi_2O_7$ thin films. Preprint at http://arxiv.org/abs/2505.12603 (2025).

22  Li, P. *et al.* Angle-resolved photoemission spectroscopy of superconducting $(La,Pr)_3Ni_2O_7/SrLaAlO_4$ heterostructures. *Natl. Sci. Rev.*, nwaf205 (2025).

23  Wang, B. Y. *et al.* Electronic structure of compressively strained thin film $La_2PrNi_2O_7$. Preprint at http://arxiv.org/abs/2504.16372 (2025).

24  Shen, J. *et al.* Anomalous energy gap in superconducting $La_{2.85}Pr_{0.15}Ni_2O_7/SrLaAlO_4$ heterostructures. Preprint at http://arxiv.org/abs/2502.17831 (2025).

25  Yang, J. *et al.* Orbital-dependent electron correlation in double-layer nickelate $La_3Ni_2O_7$. *Nat. Commun.* **15**, 4373 (2024).

26  Sun, W. *et al.* Electronic and transport properties in Ruddlesden-Popper neodymium nickelates $Nd_{n+1}Ni_nO_{3n+1}$ ($n = 1–5$). *Phys. Rev. B* **104**, 184518 (2021).

27  Sun, W. *et al.* Electronic structure of superconducting infinite-layer lanthanum nickelates. *Sci. Adv.* **11**, eadr5116 (2025).

28  Lanzara, A. *et al.* Evidence for ubiquitous strong electron–phonon coupling in high-temperature superconductors. *Nature* **412**, 510-514 (2001).

29  Kaminski, A. *et al.* Renormalization of spectral line shape and dispersion below $T_c$ in $Bi_2Sr_2CaCu_2O_{8+\delta}$. *Phys. Rev. Lett.* **86**, 1070-1073 (2001).

30  Zhang, Y., Lin, L.-F., Moreo, A. & Dagotto, E. Electronic structure, dimer physics, orbital-selective behavior, and magnetic tendencies in the bilayer nickelate superconductor $La_3Ni_2O_7$ under pressure. *Phys. Rev. B* **108**, L180510 (2023).

31  Bhatta, H. C. R. B., Zhang, X., Zhong, Y. & Jia, C. Structural and electronic evolution of bilayer nickelates under biaxial strain. Preprint at http://arxiv.org/abs/2502.01624 (2025).

32  Geisler, B., Hamlin, J. J., Stewart, G. R., Hennig, R. G. & Hirschfeld, P. J. Fermi surface reconstruction and enhanced spin fluctuations in strained $La_3Ni_2O_7$ on $LaAlO_3$ (001) and $SrTiO_3$ (001). Preprint at http://arxiv.org/abs/2411.14600 (2025).

33  Zhao, Y.-F. & Botana, A. S. Electronic structure of Ruddlesden-Popper nickelates: Strain to mimic the effects of pressure. *Phys. Rev. B* **111**, 115154 (2025).

34  Wang, H. *et al.* Electronic structures across the superconductor-insulator transition at $La_{2.85}Pr_{0.15}Ni_2O_7/SrLaAlO_4$ Interfaces. Preprint at http://arxiv.org/abs/2502.18068 (2025).

35  Sato, T., Kamiyama, T., Takahashi, T., Kurahashi, K. & Yamada, K. Observation of $d_{x^2-y^2}$-like superconducting gap in an electron-doped high-temperature superconductor. *Science* **291**, 1517-1519 (2001).

36  Xu, K.-J. *et al.* Bogoliubov quasiparticle on the gossamer Fermi surface in electron-doped cuprates. *Nat. Phys.* **19**, 1834-1840 (2023).

37  Sobota, J. A., He, Y. & Shen, Z.-X. Angle-resolved photoemission studies of quantum materials. *Rev. Mod. Phys.* **93**, 025006 (2021).

**Acknowledgements**

This work was supported by the National Key R&D Program of China (Grant Nos. 2021YFA1400400, 2022YFA1402502, 2023YFA1406304 and 2024YFA1408103), National Natural Science Foundation of China (Grant Nos. 12434002, 123B2051 12494593, and 12204394), Natural Science Foundation of Jiangsu Province (Grant No. BK20233001), Natural Science Foundation of Anhui Province (Grant No. 2408085J003), Fundamental Research Funds for the Central Universities (Grant Nos. 021314380269, 021314380277), New Cornerstone Science Foundation, and the Innovation Program for Quantum Science and Technology (Grant No. 2021ZD0302803). D.J. acknowledges the Hong Kong Research Grants Council General Research Fund (Grant Nos. 15303923 and 15307224). Z.J. acknowledges the Postdoctoral Fellowship Program and China Postdoctoral Science Foundation (Grant No. BX20240348). H.S. acknowledges the China National Postdoctoral Program for Innovative Talents (Grant No. BX20230152), the China Postdoctoral Science Foundation (Grant No. 2024M751368). Part of this research used Beamline 03U of the Shanghai Synchrotron Radiation Facility, which is supported by $ME^2$ project under Contract No.11227902 from National Natural Science Foundation of China. The numerical calculations in this paper have been done on the computing facilities in the High-Performance Computing Center (HPCC) of Nanjing University.

**Author contributions**

Y.N. conceived the project and directed the project with D.S. and D.F. W.S., Z.J., H.Z. and Z.L. conducted the ARPES measurement at Shanghai Synchrotron Radiation Facility. W.S., B.H., S.Y., and M.W. synthesized nickelate films and performed structural characterizations and transport measurements. Y.Y. and D.J. performed the STEM measurements. J.Z. performed DFT calculations. W.S., Z.J., B.H., H.S. Z.G., D.S., D.F. and Y.N. wrote the manuscript with input from all authors.

**Competing interests**

The authors declare no conflict of interest.

**Data availability**

The data that support the findings of this study are available from the corresponding author on reasonable request.

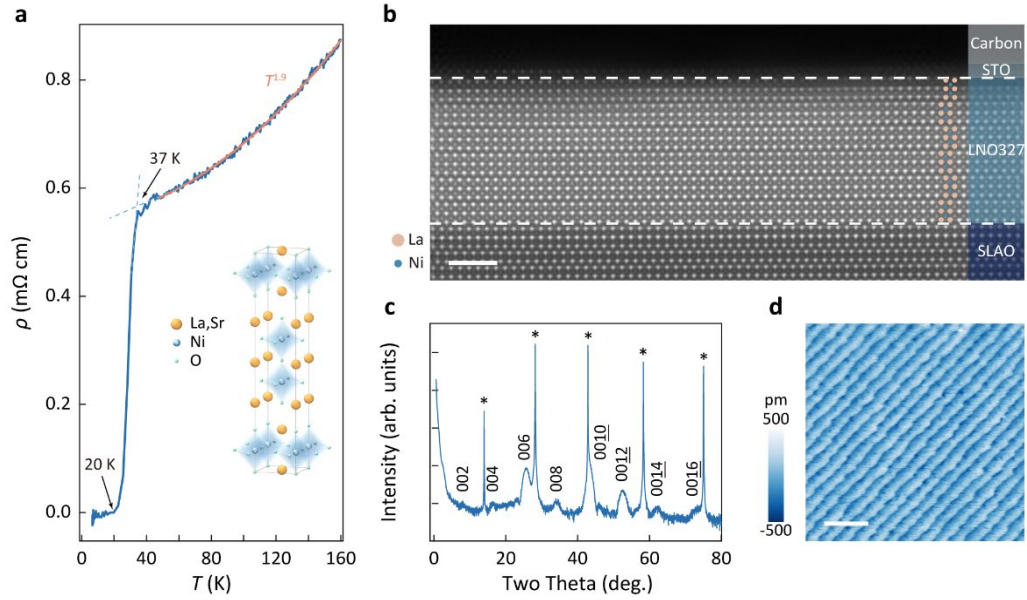

**Fig. 1 | Structural and transport characterizations of *in-situ* ozone-treated La$_{2.79}$Sr$_{0.21}$Ni$_2$O$_7$ thin film**. **a**, Temperature-dependent resistivity of a representative 2-unit-cell (u.c.)-thick La$_{2.79}$Sr$_{0.21}$Ni$_2$O$_7$ thin film, showing the superconducting onset and zero-resistance temperature of 37 and 20 K, respectively. Inset shows a schematic illustration of a single unit cell in the La$_{2.79}$Sr$_{0.21}$Ni$_2$O$_7$ Ruddlesden-Popper (RP) structure. **b**, Cross-sectional high-angle annular dark-filed (HAADF) scanning transmission electron microscopy (STEM) image of a representative 3-u.c.-thick La$_{2.79}$Sr$_{0.21}$Ni$_2$O$_7$ thin film. The alternating stacking of [NiO$_2$] bilayers is clearly observed, with no detectable intergrowths of other type of Ruddlesden-Popper (RP) phases. One-u.c.-thick SrTiO$_3$ capping layer was deposited on the top of the La$_{2.79}$Sr$_{0.21}$Ni$_2$O$_7$ thin film for protection during focused ion beam (FIB) lamella preparation. **c**, X-ray diffraction 2*θ*-*ω* scans of a representative 2-u.c.-thick La$_{2.79}$Sr$_{0.21}$Ni$_2$O$_7$ thin film, exhibiting phase-pure (00*l*)-oriented diffraction peaks. The diffraction peaks from SrLaAlO$_4$ substrate are marked by asterisks. **d**, Surface topography of *in-situ* ozone-treated La$_{2.79}$Sr$_{0.21}$Ni$_2$O$_7$ thin film measured by atomic force microscopy. Scale bars, 2 nm (**b**); 2 *μ*m (**d**).

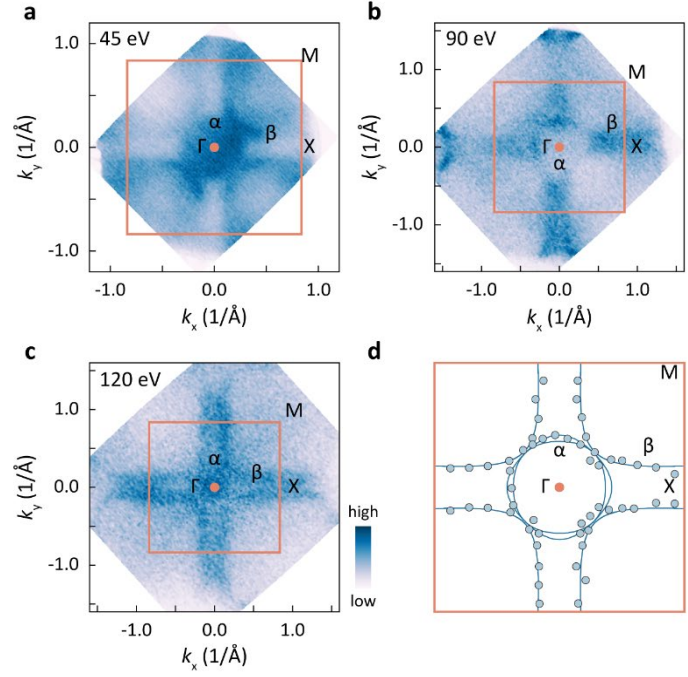

**Fig. 2 | Fermi surface of La$_{2.79}$Sr$_{0.21}$Ni$_2$O$_7$ thin film**. **a-c**, Fermi surface (FS) mappings measured at 7 K with 45, 90 and 120 eV photons, respectively. Orange boxes indicate the first Brillouin zone (BZ). The energy integration window is ±15 meV symmetrically centered at the Fermi level ($E_F$). **d**, Fermi momenta (filled circles) for $\beta$ band extracted from FS mapping in **a** and **c**, and for $\alpha$ band extracted from FS mapping in **b.** The calculated FS (blue solid lines) under 2% compressive strain using the density functional theory is overlaid for comparison.

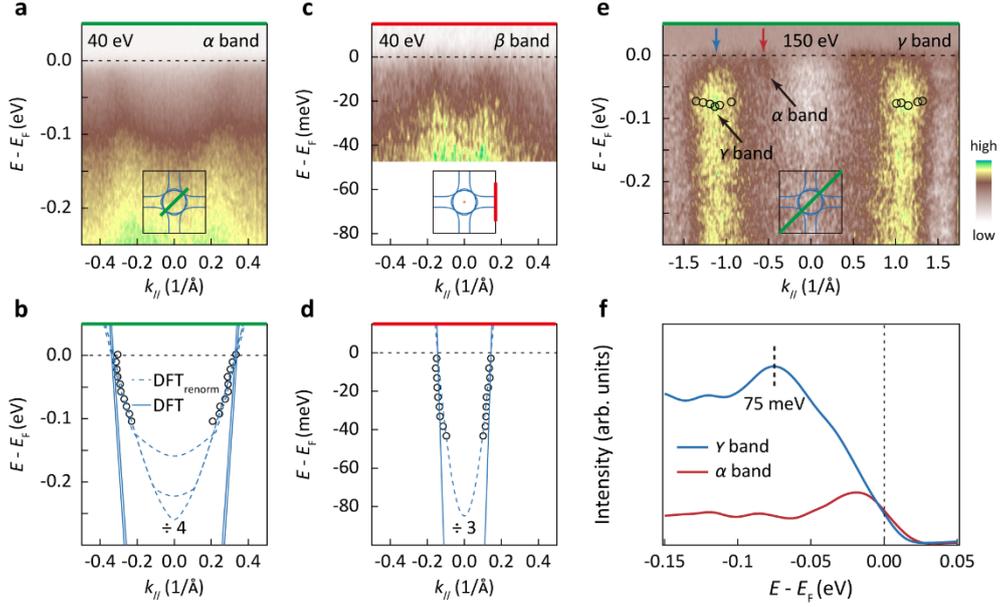

**Fig. 3 | Band dispersions and renormalizations in La$_{2.79}$Sr$_{0.21}$Ni$_2$O$_7$ thin film**. **a**, Band structure measured at 7 K with 40 eV photons along the diagonal direction of the BZ (green solid line in the inset). **b**, Band dispersions extracted from the peak positions of momentum distribution curves (MDC) in **a**. The DFT bare bands ($U = 0$) and corresponding renormalized bands are shown in solid and dashed lines, respectively. **c-d**, same as **a-b**, but measured along the BZ boundary. **e**, Background-subtracted spectrum measured at 7 K with 150 eV photons along the diagonal direction of the BZ (green solid line in the inset). Black circles represent peak positions of corresponding energy distribution curves (EDCs). The background is generated by averaging EDCs at the momentum range of 1.5-1.7 1/Å where low energy bands are absent (Extended Data Fig. 3). **f**. Smoothed background-subtracted EDCs for $\gamma$ and $\alpha$ bands taken at momentum region indicated by the blue and red arrows in **e**, respectively.

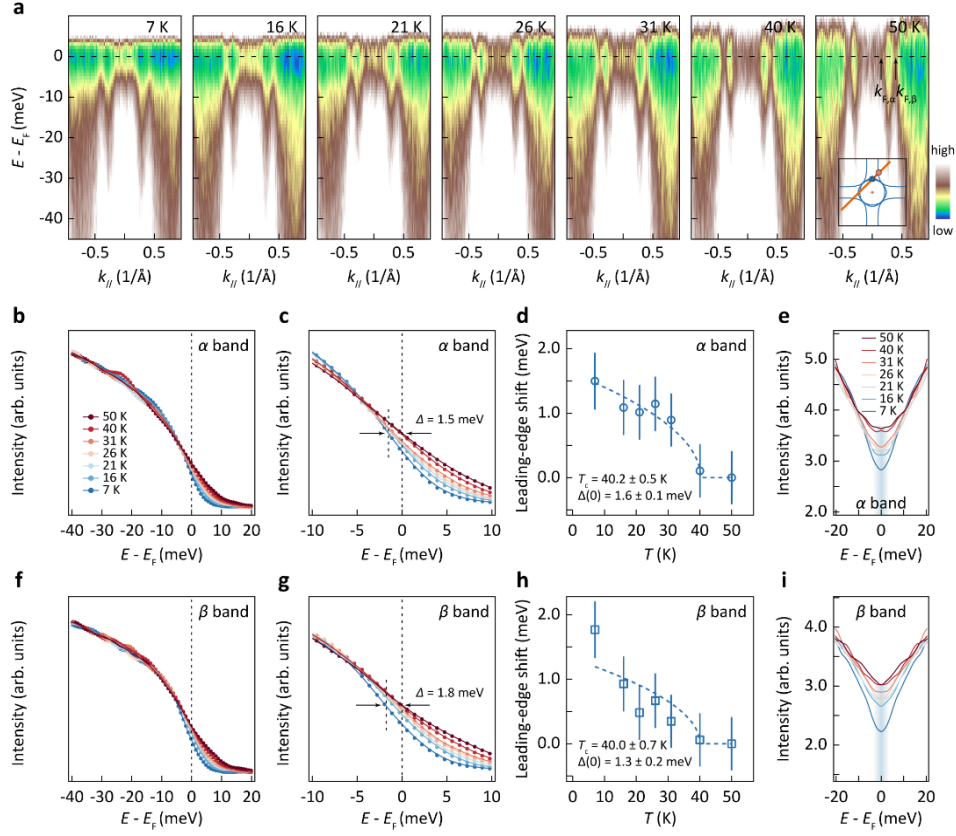

**Fig. 4 | Leading-edge shifts near the diagonal direction across the superconducting transition in La$_{2.79}$Sr$_{0.21}$Ni$_2$O$_7$ thin film**. **a**, ARPES spectra taken near the diagonal direction (orange solid line in the inset) with 45 eV photons at different temperatures. All spectra are divided by the Fermi-Dirac distributions to reflect the particle-hole symmetric feature near the $E_F$. **b**, Normalized EDCs taken at the Fermi momentum of the α band ($k_{F,α}$) as indicated by the black arrow and blue dot (inset BZ) in **a**. Solid circles are the experimental data, whereas solid lines are the smoothed average to them. All EDCs are normalized to the area between binding energies of 30–40 meV. **c**, Zoomed-in view near $E_F$ of **b**. Leading-edge shifts of EDC for 7 K are indicated by the double arrows. **d**, Extracted leading-edge shifts from **c**. The dashed line corresponds to the theoretical fit employing the weak-coupling BCS energy-gap function. Error bars represent uncertainties coming from fitting procedure and Fermi energy calibration. **e**, Normalized symmetrized EDCs for α band, with the spectral-weight change near $E_F$ highlighted (blue shaded area). All EDCs are normalized to the area between binding energies of 30–40 meV. **f-i**, same as **b-e**, but corresponding to the β band.

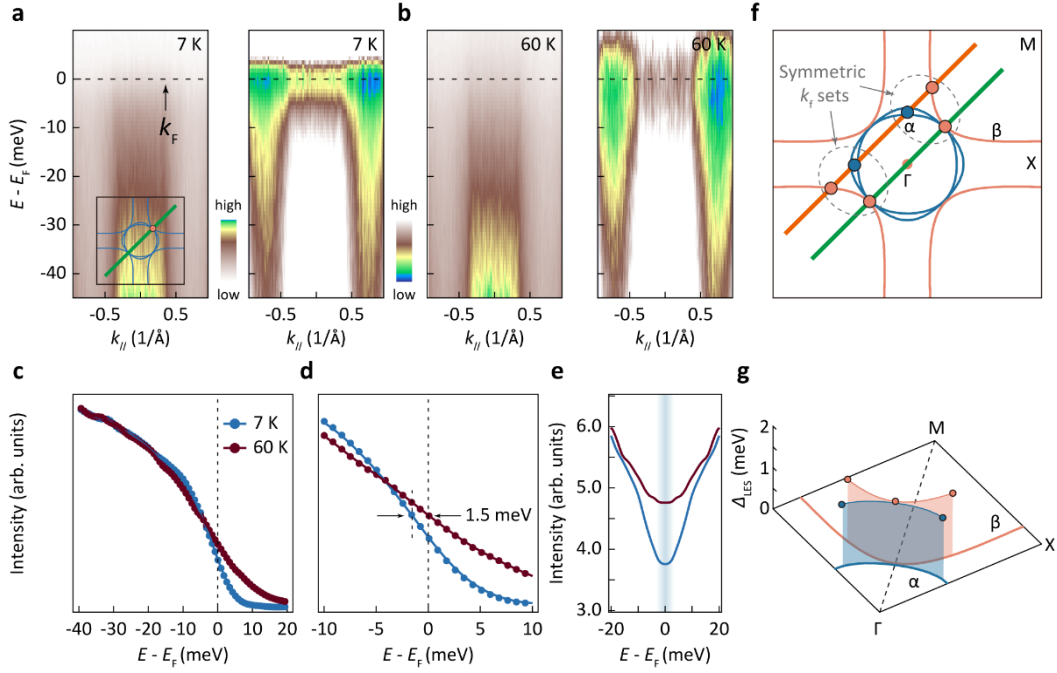

**Fig. 5 | Leading-edge shifts at the diagonal direction across the superconducting transition in La$_{2.79}$Sr$_{0.21}$Ni$_2$O$_7$ thin film. a,** Raw ARPES spectrum (left column) taken at the diagonal direction of the first BZ (green solid line in the inset) with 45 eV photons at 7 K, and corresponding Fermi-Dirac-divided spectrum (right column). **b,** same as **a**, but taken at 60 K. **c,** Normalized EDCs taken at the Fermi momentum ($k_F$) as indicated by the black arrow and orange dot (inset BZ) in **a**. Solid circles are the experimental data, whereas solid lines are the smoothed average to them. All EDCs are normalized to the area between binding energies of 30–40 meV. **d,** Zoomed-in view near $E_F$ of **c**. Leading-edge shifts of EDC for 7 K are indicated by the double arrows. **e,** Normalized symmetrized EDCs at the $k_F$. It can be seen that the spectral weight near $E_F$ is depleted with decreasing temperatures. All EDCs are normalized to the area between binding energies of 30–40 meV. **f,** The calculated FS (blue and orange solid lines) under 2% compressive strain, overlaid with two momentum cuts along which the measurements were conducted. **g,** Schematic illustrations of the momentum-dependent leading-edge shifts ($\Delta_{LES}$).

18